\documentclass[sigconf]{acmart}

\usepackage[english]{babel}
\usepackage{blindtext}
\usepackage{tikz}
\usepackage{amsmath}
\usepackage{tabularx}
\usepackage{graphicx}
\usepackage{subcaption}
\usepackage{multirow}
\usepackage{array}
\usepackage{booktabs}
\usepackage{tikz}
\usepackage{algorithm}
\usepackage{algpseudocode}
\usepackage{makecell}
\usepackage{xcolor,colortbl}
\usepackage{siunitx}
\usepackage{xspace}
\usepackage[title]{appendix}

\usetikzlibrary{shapes.geometric, shapes.multipart, arrows, positioning}
\newcolumntype{C}[1]{>{\centering\arraybackslash}m{#1}}

\definecolor{F1DeepGreen}{HTML}{52b788}
\definecolor{F1MediumGreen}{HTML}{74c69d}
\definecolor{F1LightGreen}{HTML}{95d5b2}
\definecolor{F1LighterGreen}{HTML}{b7e4c7}
\definecolor{F1LightestGreen}{HTML}{d8f3dc}

\definecolor{TimeDeepBlue}{HTML}{00b4d8}
\definecolor{TimeMediumBlue}{HTML}{48cae4}
\definecolor{TimeLightBlue}{HTML}{90e0ef}
\definecolor{TimeLighterBlue}{HTML}{ade8f4}
\definecolor{TimeLightestBlue}{HTML}{caf0f8}

\renewcommand{\paragraph}[1]{\vspace*{0.03in}\noindent\textbf{#1}}

\renewcommand\footnotetextcopyrightpermission[1]{} % removes footnote with conference info
\setcopyright{none}
\acmDOI{}

\acmISBN{}

\acmConference[Working Draft]{}
\acmYear{\today}
\acmPrice{}

\newcommand{\copper}{\textit{(10) Copper Wire}\xspace}

\newcommand{\cable}{\textit{(40) Coaxial Cable - HFC}\xspace}

\newcommand{\fiber}{\textit{(50) Optical Carrier - Fiber to the Premises}\xspace}

\newcommand{\geosat}{\textit{(60) Geostationary Satellite}\xspace}

\newcommand{\ngssat}{\textit{(61) Non-geostationary Satellite}\xspace}

\newcommand{\utfwl}{\textit{(70) Unlicensed Terrestrial Fixed Wireless}\xspace}

\newcommand{\ltfwl}{\textit{(71) Licensed Terrestrial Fixed Wireless}\xspace}
\newcommand{\ltfwlabbr}{\textit{(71) LTF Wireless}\xspace}

\newcommand{\lbrtfwl}{\textit{(72) Licensed-by-Rule Terrestrial Fixed Wireless}\xspace}

\newcommand{\noapptabbr}{\textit{(1) No Appt}\xspace}

\newcommand{\noshowabbr}{\textit{(2) No Show}\xspace}

\newcommand{\extrafeeabbr}{\textit{(3) Extra Fee}\xspace}

\newcommand{\servicedeniedabbr}{\textit{(4) Service Denied}\xspace}

\newcommand{\notechnologyabbr}{\textit{(5) No Technology}\xspace}

\newcommand{\nospeedabbr}{\textit{(6) No Speed}\xspace}

\newcommand{\nosignalabbr}{\textit{(8) No Signal}\xspace}

\begin{document}
% \title{No One Challenges Like Nebraska: An Analysis of the US Broadband Data Collection Fixed Availability Challenges}
\title[Are We Up to the Challenge?]{Are We Up to the Challenge? An analysis of the FCC Broadband Data Collection Fixed Internet Availability Challenges}

%\titlenote{Produces the permission block, and copyright information}

% \author{Paper \# 488, 12 pages body, 16 pages total}

\author{Jonatas Marques}
% \authornote{Note}
\email{jmarques@uchicago.edu}
\affiliation{%
  \institution{University of Chicago}
%   \streetaddress{USA}
%   \city{City} 
  \state{USA}
%   \postcode{Zipcode}
}

\author{Alexis Schrubbe}
\email{schrubbe@uchicago.edu}
\affiliation{%
  \institution{University of Chicago}
  \state{USA}
}

\author{Nicole P. Marwell}
\email{nmarwell@uchicago.edu}
\affiliation{%
  \institution{University of Chicago}
  \state{USA}
}

\author{Nick Feamster}
\email{feamster@uchicago.edu}
\affiliation{%
  \institution{University of Chicago}
  \state{USA}
}

\renewcommand{\shortauthors}{Marques .et al.}

\begin{abstract}
    In 2021, the Broadband Equity, Access, and Deployment (BEAD) program allocated \$42.45 billion to enhance high-speed internet access across the United States.
    As part of this funding initiative, The Federal Communications Commission (FCC) developed a national coverage map to guide the allocation of BEAD funds.
    This map was the key determinant to direct BEAD investments to areas in need of broadband infrastructure improvements.
    The FCC encouraged public participation in refining this coverage map through the submission of “challenges” to either locations on the map or the status of broadband at any location on the map.
    These challenges allowed citizens and organizations to report discrepancies between the map's data and actual broadband availability, ensuring a more equitable distribution of funds.
    In this paper, we present a study analyzing the nature and distribution of these challenges across different access technologies and geographic areas.
    Among several other insights, we observe, for example, that the majority of challenges (about 58\%) were submitted against terrestrial fixed wireless technologies as well as that the state of Nebraska had the strongest engagement in the challenge process with more than 75\% of its broadband-serviceable locations having submitted at least one challenge.
\end{abstract}

\maketitle

\section{Introduction} \label{sec:introduction}

Access to the Internet—much like energy, water, and sanitation—is essential.
People rely on the Internet for work, entertainment, education, health care, and more.
Despite the Internet's prominence in the US economy, barriers in access and adoption still persist across the nation.
In response to persistent gaps in the diffusion of high-speed, reliable broadband, and to help close gaps in access and achieve the goal of universal service across the United States (US), the US Federal Government allocated significant investments for the expansion of broadband as part of the Infrastructure Investment and Jobs Act (IIJA) \cite{IIJA}.
The Broadband Equity, Access, and Deployment (BEAD) program, as it is called, provides more than 42 billion dollars to expand high-speed Internet access through planning, infrastructure deployment, and adoption programs across the country \cite{BEADProgramLaunch, BEAD}.

In order to determine how to distribute BEAD funds among US states and territories, the Federal Communications Commission (FCC) was assigned to collect and visualize data through the Broadband Data Collection (BDC) program, which produced the National Broadband Map (NBM) \cite{NBM}.
This map compiles two products: the location fabric which is an underlying fabric of broadband serviceable locations (BSLs) and the availability data layered on top of the fabric.
The availability data is compiled from Internet Service Providers (ISPs) regarding their broadband offerings to depict fixed Internet access availability for every location in the country \cite{whatsonNBM}.
Depending on the maximum available speeds offered in a particular BSL, that BSL considered either being unserved (no access to 25/3 Mbps for download/upload), underserved (no access to 100/20 Mbps), or served (access to 100/20 Mbps) \cite{BEAD}.
The status of each BSL, as shown in the map, determines its eligibility to receive funds from the BEAD program.
Both unserved and underserved are considered eligible, with preference given to fund projects to connect the former.
Served locations are considered ineligible for BEAD funding.

This context highlights the importance that the FCC's BDC datasets contain data that accurately represents the lived-reality of broadband consumers across the US.
Consequently, as part of this project, the FCC has made an historic call for engagement from citizens to make the broadband map accurate \cite{BEADChallengeProcess}.
Citizens and organizations were tasked with checking the reported availability at their familiar locations (e.g., households, workplaces) and raising challenges refuting the status of 1) BSL locations in the fabric or 2) availability at any given BSL for every instance where the BDC data did not match their lived experience.
The majority of challenges to the BDC data occurred during the opening of the Federal Availability Challenge Process which took place between November 18th 2022 and January 13th 2023.
The outcome of Federal Availability Challenge Process determined the proportion of BEAD funding each state and territory would receive.
Since that process the FCC has regularly published datasets reporting on submitted challenges (e.g, their associated locations, providers, access technologies, and reasons for challenge) as well as their outcomes (i.e., upheld, overturned, or withdrawn) \cite{BDCDataSpec}.
Challenges to the map after the close of January 13th, 2023 did not impact the distribution of funds among states, but serve to make the FCC maps more accurate and may have effects into the distribution of funds within states.

In this paper, we present our analysis of the BDC challenges.
We also present several contributions that may give insight into different states' engagement with and consequent success or challenge participating in this process.
The purpose of this exploration is to present our main findings regarding the distribution of challenges across different geographic levels.
We raise interesting observations on how different states and counties have engaged with the challenge process, where they were successful and where they were not, where they focused on specific types of access technology, submitted challenges due to specific reasons, and more.
% We also build statistical models based on demographics variables that are capable of explaining X\% of the variability in number and type of challenges across geography units.

\section{Dataset} \label{sec:dataset}

In this study we mainly focus on the analysis of Fixed Challenge data regularly published as part of the FCC Broadband Data Collection (BDC) program \cite{BDCDataSpec}.
We consider all data since the first publication (November, 2022) until the latest publication available to date (January, 2024), sixteen months in total.
This dataset contains information on challenges submitted seeking to correct inaccuracies present in the Availability dataset published by the same program.
The Availability dataset describes, for each broadband-serviceable location (BSL), which Internet service providers (ISPs) offer Internet service at what speeds and using what access technologies.

\paragraph{Data Specification.} The Fixed Challenge dataset contains a number of fields that uniquely identify each challenge and describe the involved BSL and ISP, the reason for challenge submission, and the outcome.
Next, we summarize the main fields of interest for this study.

\begin{itemize}
    \item \textit{challenge\_id}: A unique identifier for the challenge.

    \item \textit{location\_id}: A unique identifier for the broadband serviceable location (BSL) associated to the challenge. Each challenge is associated to a single BSL.
    
    \item \textit{provider\_id}: A unique identifier for the Internet service provider (ISP) being challenged. Each challenge is associated to a single ISP.
    
    \item \textit{technology}: A code that identifies the access technology used for the service subject to challenge. Each challenge is associated to a single technology. Possible values are shown in Table~\ref{tab:technologies}.
    
    \item \textit{category\_code}: A code that identifies the category reason for the challenge being raised. Each challenge is associated to a single category. Possible values are shown in Table~\ref{tab:categories}.
    
    \item \textit{outcome}: The status of the challenge after either a provider concession, challenger's withdrawal, or FCC adjudication. Possible outcome values are: upheld, overturned, or withdrawn.
\end{itemize}

\begin{table}[thbp]
    \centering
    % \footnotesize
    \begin{tabular}{cp{4cm}c}
    \toprule
    \textbf{Code} & \textbf{Technology} & \textbf{Abbreviation}\\
    \midrule
    10 & Copper Wire & Copper \\
    40 & Coaxial Cable / HFC & Cable \\
    50 & Optical Carrier / Fiber to the Premises & Fiber \\
    60 & Geostationary Satellite & GS Sat \\
    61 & Non-geostationary Satellite & NGS Sat \\
    70 & Unlicensed Terrestrial Fixed Wireless & UTF Wireless \\
    71 & Licensed Terrestrial Fixed Wireless & LTF Wireless \\
    72 & Licensed-by-Rule Terrestrial Fixed Wireless & LbRTF Wireless \\
    \bottomrule
    \end{tabular}
    \caption{Access Technology Codes and Descriptions.}
    \label{tab:technologies}
\end{table}

\begin{table}[thbp]
    \centering
    \begin{tabular}{cp{4cm}c}
    \toprule
    \textbf{Code} & \textbf{Category} & \textbf{Abbreviation}\\
    \midrule
    1 & Provider failed to schedule a service installation within 10 business days of request. & No Appt \\
    2 & Provider did not install the service at the agreed-upon time. & No Show \\
    3 & Provider requested more than the standard installation fee to connect service. & Extra Fee \\
    4 & Provider denied a request for service. & Service Denied \\
    5 & Provider does not offer the technology reported to be available at this location. & No Technology \\
    6 & Provider does not offer the speed(s) reported to be available at this location. & No Speed \\
    8 & A wireless or satellite signal is not available at this location. & No Signal \\
    9 & Provider needed to construct new equipment at this location. & New Equipment \\
    \bottomrule
    \end{tabular}
    \caption{Challenge Category Codes and Descriptions.}
    \label{tab:categories}
\end{table}

\paragraph{Data Wrangling.} In the process of preparing the publicly available raw data for easy access and analysis, we carried out a few data cleaning and transformation steps.
First, we noticed that about a thousand \textit{challenge\_id} values appeared in the dataset more than once.
We determined that this is due to the fact that new challenge data is released at the end of every month.
Each new release may update the outcome of some challenges reported in previous releases due to, for example, appeals on the FCC adjudication.
To treat these cases, whenever a \textit{challenge\_id} appears more than once in the dataset, we only keep the most recently reported record, which reflects the final outcome of the challenge to date.
Second, for our study, which involves spatial analysis, it is also important to know the geographical location of each BSL involved in a challenge.
This information is not readily available in the Fixed Challenge dataset.
Nevertheless, the Availability dataset contains the field \textit{block\_geoid}, which indicates the 15-digit U.S. Census Bureau FIPS code for the census block in which each BSL is located.
This location information is sufficiently fine-grained for the purposes of our study.
Given that, we determine the location of each challenging BSL by performing a left join on the \textit{location\_id} field between the Fixed Challenge and Availability datasets.

\section{Descriptive Statistics} \label{sec:descriptive-stats}

This section describes descriptive statistical analyses conducted focusing on challenge distributions across geographic units at different levels, technologies, categories and outcomes. Each following section presents the results obtained and observations made at each evaluated geographic level.

\subsection{Nationwide Analysis}

We begin our National analysis by looking at overall statistics considering all challenges present in the dataset.
A total of 3,690,772 challenges were submitted.
These challenges were distributed among 2,872,304 unique BSLs.
We refer to these BSLs, those that have at least one associated challenge, as engaged BSLs.
The set of engaged BSLs represents about 2.4\% of all the BSLs present in the availability dataset (119,446,162).
These challenges targeted availability reports from 872 distinct Internet service providers (ISPs).

\begin{table}[!t]
    \centering
    \includegraphics[scale=0.18]{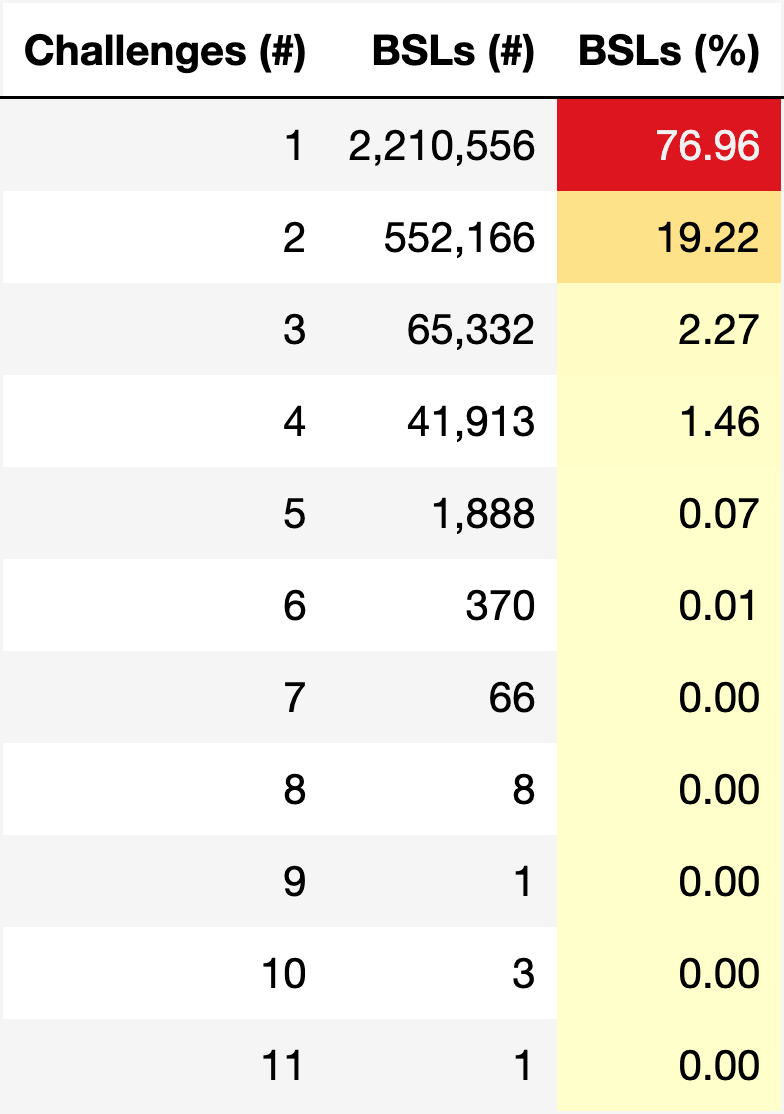}
    \caption{Number and percentage of Broadband-Serviceable Locations (BSLs) given the number of submitted challenges.}
    \label{tab:bsls_per_n_challenges}
\end{table}

\paragraph{Distribution of challenges among engaged BSLs.}
We counted the number of challenges submitted for each BSL and show the results in Tab.~\ref{tab:bsls_per_n_challenges}.
Each row indicates for a specific number of challenges (\textit{Challenges (\#)}) how many (\textit{BSLs (\#)}) and what percentage (\textit{BSLs (\%)}) of BSLs had exactly that number of challenges submitted.
The challenges appear widely spread among engaged BSLs.
About 77\% of these BSLs had only a single associated challenge.
Further, about 96\% had two challenges or fewer and about 99.9\% had four challenges or fewer.
Finally, the maximum number of challenges associated to a single BSL was eleven.

\begin{table}[!t]
    \centering
    \includegraphics[scale=0.18]{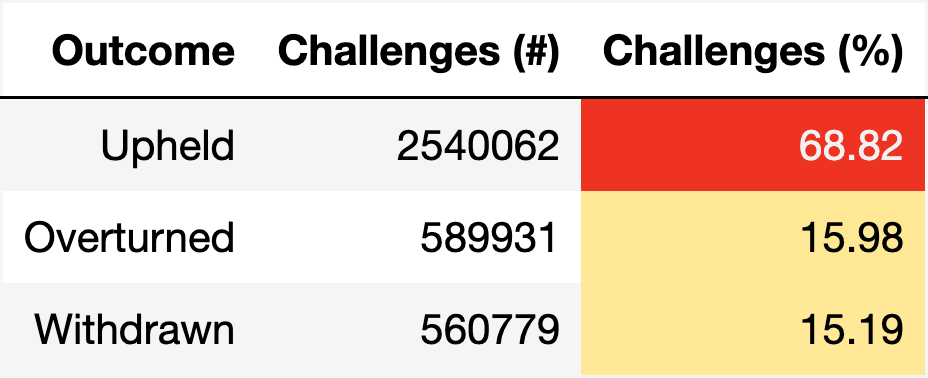}
    \caption{Distribution of challenges across outcomes.}
    \label{tab:challenges_per_outcome}
\end{table}

\paragraph{Distribution of challenge outcomes.}
We also counted the number and rate at which challenges were upheld, overturned, or withdrawn, as shown in Tab.~\ref{tab:challenges_per_outcome}.
The majority (about 69\%) of challenges  were \textit{upheld}, correcting two and a half million points of availability data.
About 16\% of the submitted challenges were \textit{overturned}.
Perhaps surprisingly, about 15\% of challenges were \textit{withdrawn}.
Overturned and withdrawn numbers considered together, represent more than a million availability data points that were more closely inspected by the FCC and found to be accurate.

\begin{table}[!t]
    \centering
    \includegraphics[scale=0.18]{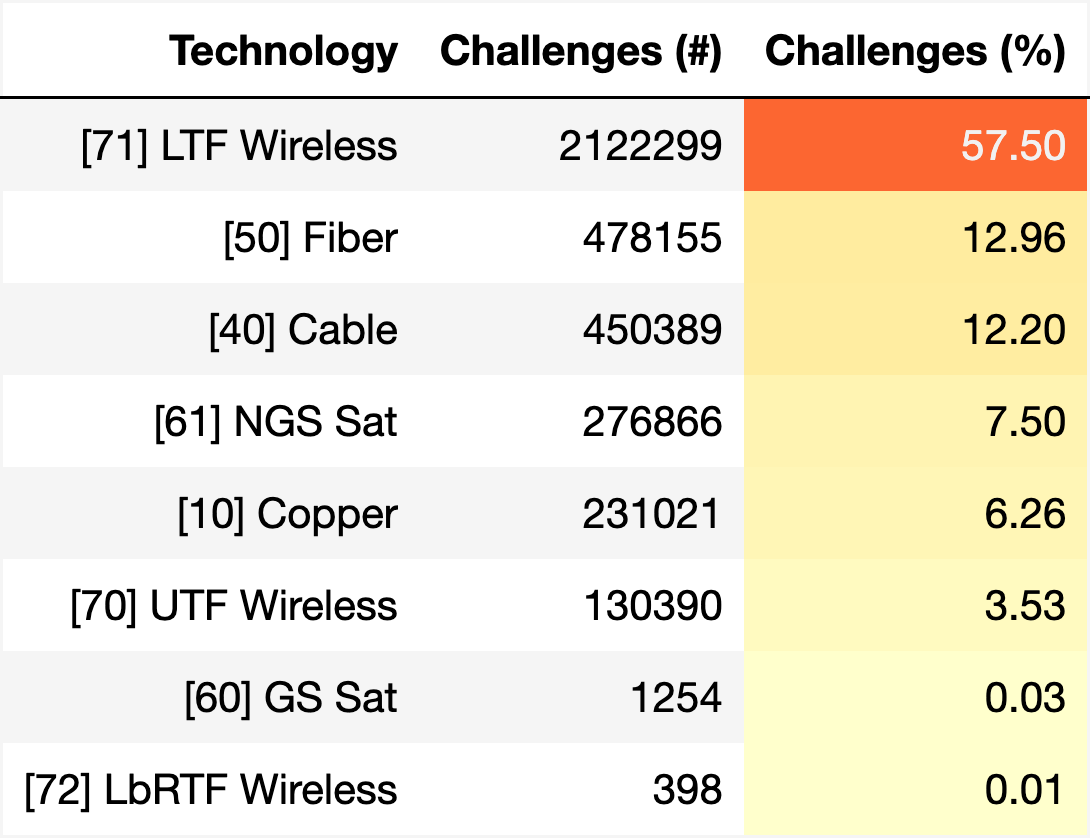}
    \caption{Distribution of challenges across technologies.}
    \label{tab:challenges_per_technology}
\end{table}

\paragraph{Distribution of challenges across technologies.}
Tab.~\ref{tab:challenges_per_technology} presents the number and rate at which challenges targeted distinct Internet access technologies.
The majority (about 58\%) of challenges were submitted against \ltfwl.
% This may suggest frequent inaccuracy in reporting the availability of this technology by ISP, possibly due to the intrinsic difficulties of determining which BSLs are covered by wireless signal in large areas.
Other notable technologies were \fiber and \cable with about 13\% and 12\% of the challenges, respectively.
Only a tiny fraction of challenges were submitted against \geosat and \textit\lbrtfwl.

\begin{table}[!t]
    \centering
    \includegraphics[scale=0.18]{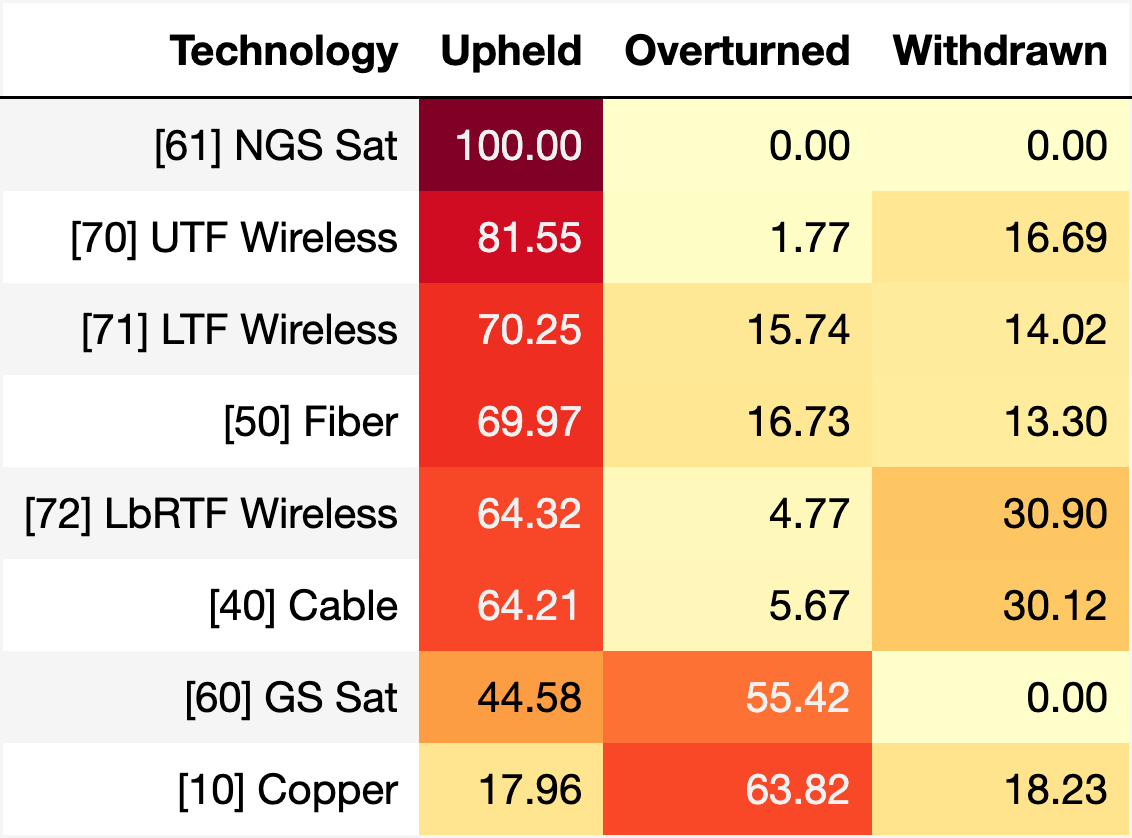}
    \caption{Distribution of outcomes across technologies. Outcome values are shown as percentages.}
    \label{tab:outcome_per_technology}
\end{table}

\paragraph{Distribution of challenge outcomes across technologies.}
Tab.~\ref{tab:outcome_per_technology} presents the rate at which challenges were upheld, overturned, or withdrawn disaggregated by technology.
For \ltfwlabbr, the most targeted technology, about \num{1490000} (or 70\%) challenges were upheld.
All (more than \num{270000}) challenges against \ngssat were upheld.
Most challenges against \copper were not upheld (about \num{189000} or 82\%).
About 64\% (or \num{147000}) of those challenges were overturned and another 18\% (or \num{42000}) were withdrawn.
Similarly, most challenges against \geosat were not upheld, although in this case they were only overturned (about 55\% or \num{695}) but not withdrawn (0\%).
Lastly, there was a significant percentage (more than 30\%) of withdrawals for \lbrtfwl and \cable.

\begin{table}[!t]
    \centering
    \includegraphics[scale=0.18]{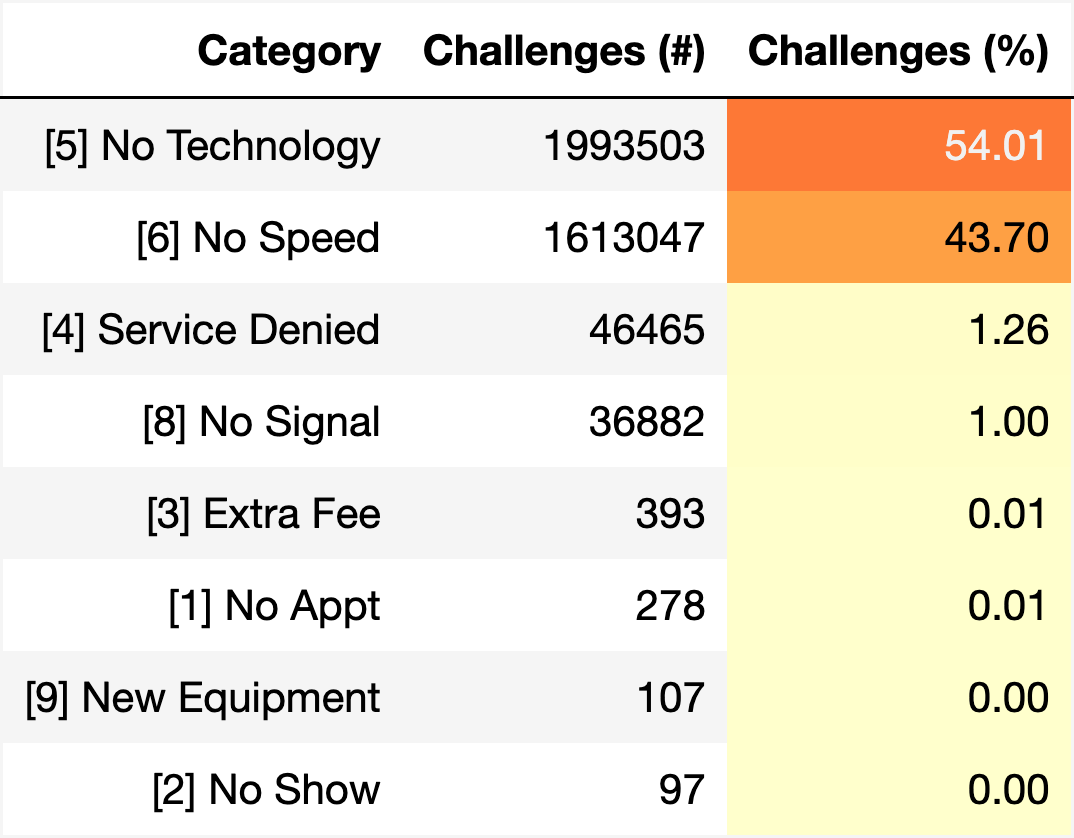}
    \caption{Distribution of challenges across categories.}
    \label{tab:challenges_per_category}
\end{table}

\paragraph{Distribution of challenges across categories.}
Tab.~\ref{tab:challenges_per_category} presents the number and rate at which challenges were submitted due to each category/reason.
The vast majority of the submitted challenges (about 98\%) argued that the provider does not offer either the technology (\notechnologyabbr, about 54\%) or the speed (\nospeedabbr, about 43\%) reported to be available at a BSL.
About another 1\% argued that the provider denied a request for service at the location (\servicedeniedabbr).
And yet another 1\% posited that a wireless or satellite signal is not available at the BSL (\nosignalabbr), contrary to provider fillings.
This overall distribution seems to make sense since the top two categories are simpler to determine, while the remaining have intrinsic timing factors or depend on a good level of interaction with the service provider before a challenge can be submitted.

\begin{table}[!t]
    \centering
    \includegraphics[scale=0.18]{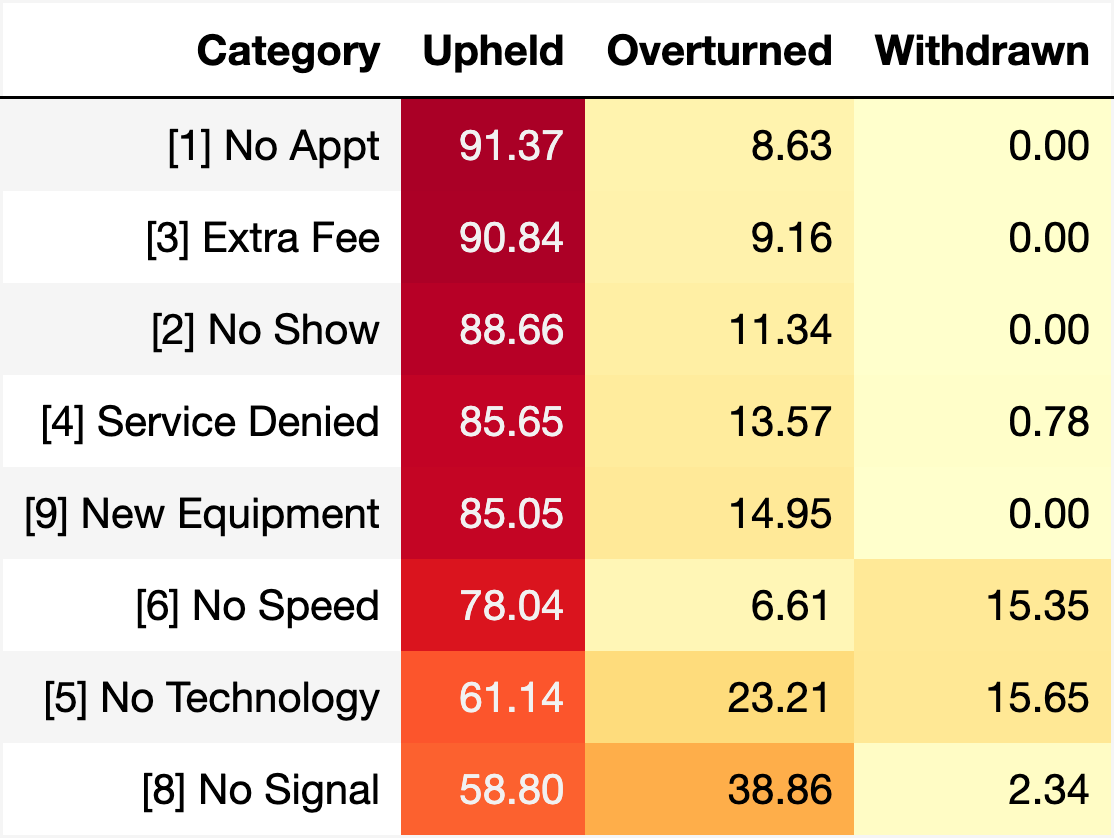}
    \caption{Distribution of outcomes across categories. Outcome values are shown as percentages.}
    \label{tab:outcome_per_category}
\end{table}

\paragraph{Distribution of challenge outcomes across categories.}
Tab.~\ref{tab:outcome_per_category}  presents the rate at which challenges were upheld, overturned, or withdrawn disaggregated by category.
Across all categories, most of the challenges were accepted.
At a minimum, about 59\% of challenges were upheld for when a wireless or satellite signal is not available at the location (\nosignalabbr).
At a maximum, about 91\% of challenges were upheld in cases where the provider failed to schedule a service installation within 10 business days of request (\noapptabbr).
Most of the withdrawn challenges argued that a provider either does not offer the technology (\notechnologyabbr) or speed (\nospeedabbr) originally reported to be available at the BSL.

\begin{table*}[!t]
    \centering
    \includegraphics[scale=0.15]{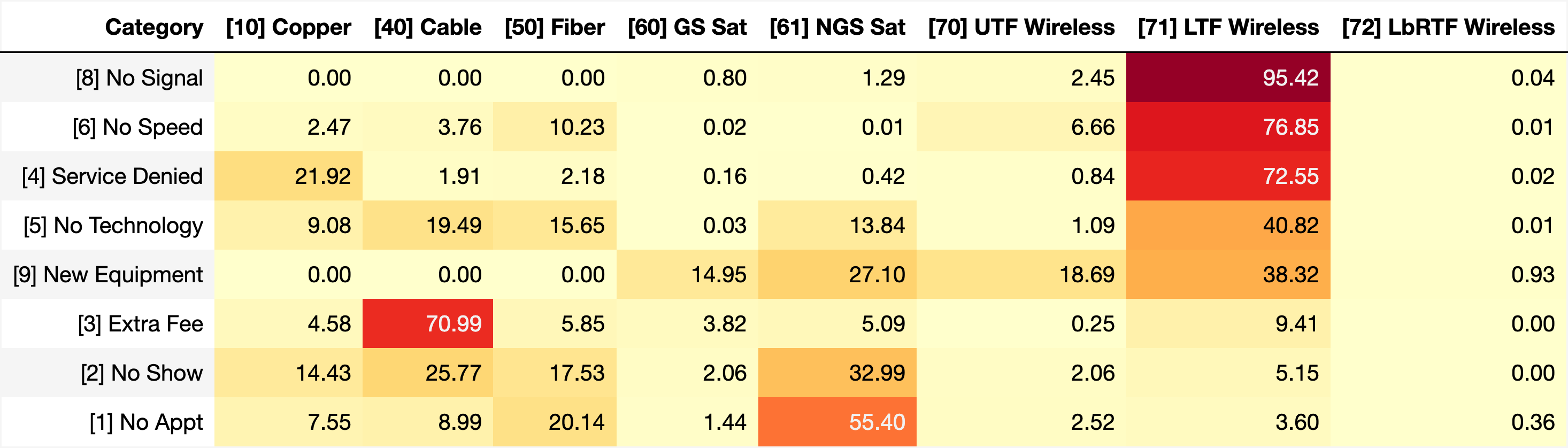}
    \caption{Distribution of challenge technologies across categories. Values are shown as percentages.}
    \label{tab:technology_per_category}
\end{table*}

\paragraph{Distribution of challenge technologies across categories.}
Tab.~\ref{tab:technology_per_category} presents the rate at which challenges were submitted against each technology disaggregated by category.
\ltfwl tops the list as the most frequent target of challenges for Categories 4, 5, 6, 8, and 9 (first five rows).
For categories \noapptabbr and \noshowabbr (last two rows), \ngssat was the top targeted technology.
For the Category \extrafeeabbr, \cable was the top targeted technology.

\begin{table*}[!t]
    \centering
    \includegraphics[scale=0.15]{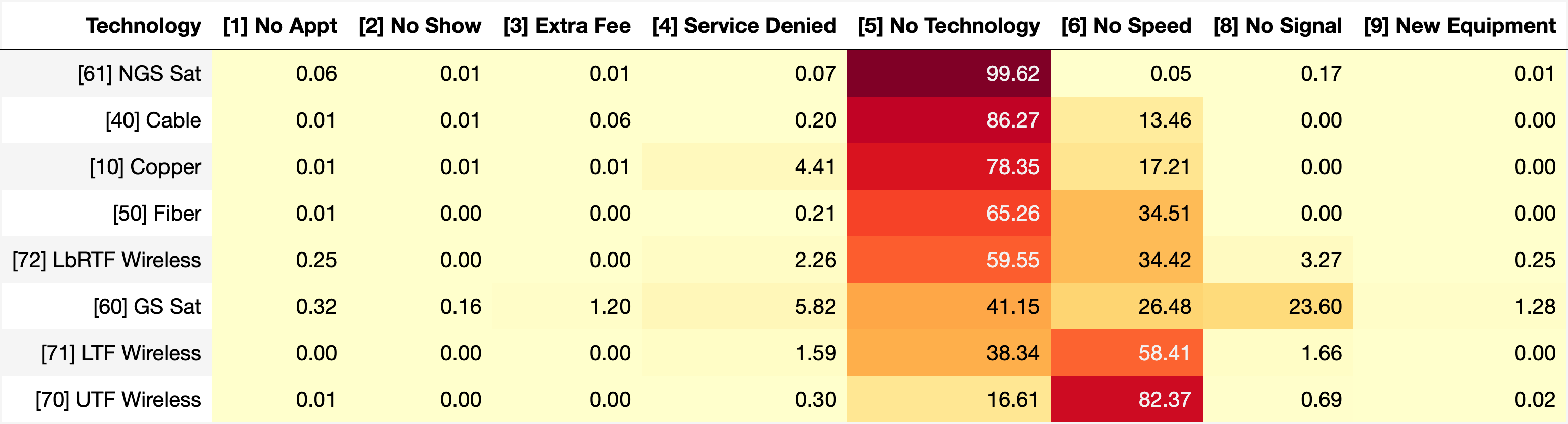}
    \caption{Distribution of challenge categories across technologies. Values are shown as percentages.}
    \label{tab:category_per_technology}
\end{table*}

\paragraph{Distribution of challenge categories across technologies.}
Tab.~\ref{tab:category_per_technology} presents the rate at which challenges were submitted due to each category/reason disaggregated by technology.
Across all technologies, categories representing that the provider does not offer the technology (\notechnologyabbr) or the speed (\nospeedabbr) reported to be available at the location usually account for the majority of the challenges.
Category \nosignalabbr, when a wireless or satellite signal is not available at the location, is always among the top-three reasons for submitting a challenge for wireless- and satellite-based technologies.

\subsection{State-level Analysis}

In this section, we analyze challenge distributions across all 50 states, Washington D.C., Puerto Rico, the U.S. Virgin Islands, Guam, American Samoa, and the Commonwealth of the Northern Mariana Islands\footnote{For simplicity, henceforth we refer to territories and DC as states.}.

%%%%%%%%%%%%%%%%%%%%%%%%%%%%%%%%%%%%%%%%%%%%%%%%%%%%%%%%%%%%%%%%%%%%%%%%%%%%%%%
% GENERAL %%%%%%%%%%%%%%%%%%%%%%%%%%%%%%%%%%%%%%%%%%%%%%%%%%%%%%%%%%%%%%%%%%%%%
%%%%%%%%%%%%%%%%%%%%%%%%%%%%%%%%%%%%%%%%%%%%%%%%%%%%%%%%%%%%%%%%%%%%%%%%%%%%%%%

\begin{table}[!t]
    \centering
    \includegraphics[scale=0.115]{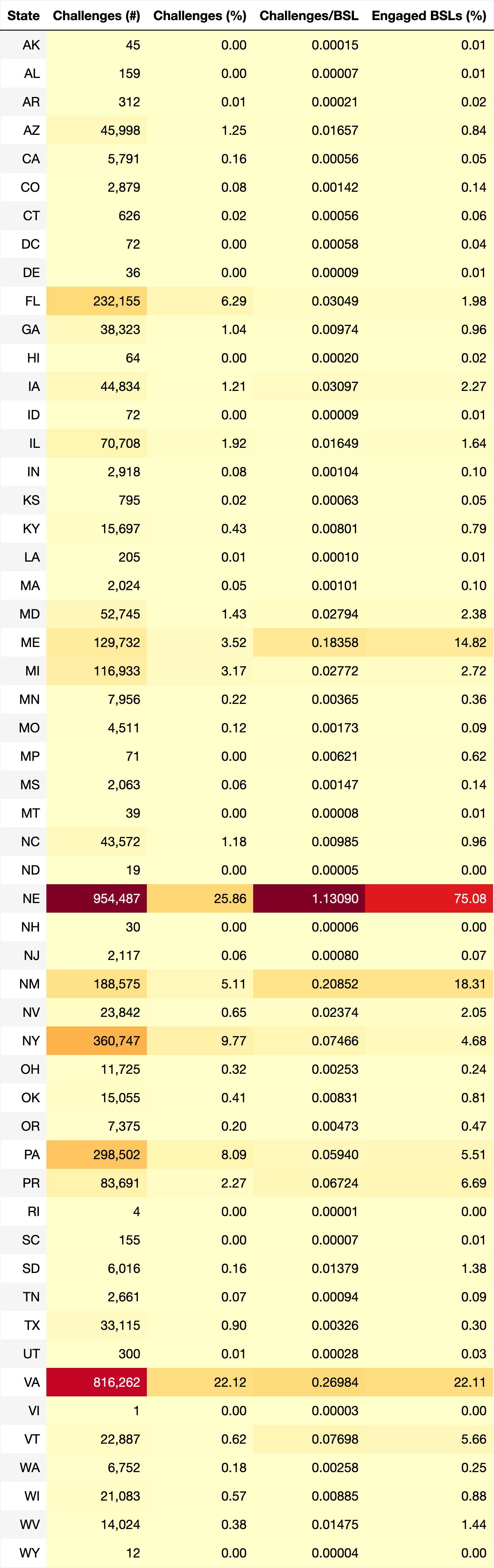}
    \caption{Big picture numbers by state.}
    \label{tab:big_picture_by_state}
\end{table}

\paragraph{Overall distribution of challenges across states.}
Table~\ref{tab:big_picture_by_state} (on Page~\pageref{tab:big_picture_by_state}) columns \textit{Challenges (\#)} and \textit{Challenges (\%)} present the absolute number of challenges submitted by each state and percentage considering all states.
From this table, we observe that challenges are not uniformly distributed across states.
Indeed, Nebraska (26\%) and Virginia (22\%) alone account for about 48\% of all the submitted challenges.
Besides these two states, six others had at least \num{100000} challenges submitted: Florida, Maine, Michigan, New Mexico, New York, and Pennsylvania.
Fifteen states had between \num{10000} and \num{100000} challenges.
Twelve had between \num{1000} and \num{10000} challenges.
Seven had between \num{100} and \num{1000} challenges.
Finally, another twelve states had less than \num{100} challenges submitted.

\paragraph{Challenge process engagement across states.}
Table~\ref{tab:big_picture_by_state} also presents the average number of challenges per BSL and the percentage of engaged BSLs in each state.
The average number of challenges per BSL is calculated as the total number of challenges divided by the number of BSLs in the state reported to have any Internet availability.
From the results, we note that even considering the number of BSLs in each state, the distribution of challenges is not uniform.
Nebraska has a unique result, it was able to engage about three quarters of its BSLs and had on average more than one challenge submitted per BSL.
Results for Virginia are not as unique as in the previous case of considering the absolute number of challenges.
Nevertheless, Virginia, Maine, and New Mexico each have about one challenge for each four or five BSLs.
Most states had low engagement, only 16 states engaged at least 1\% of its BSLs with reported Internet availability.

%%%%%%%%%%%%%%%%%%%%%%%%%%%%%%%%%%%%%%%%%%%%%%%%%%%%%%%%%%%%%%%%%%%%%%%%%%%%%%%
% OUTCOME %%%%%%%%%%%%%%%%%%%%%%%%%%%%%%%%%%%%%%%%%%%%%%%%%%%%%%%%%%%%%%%%%%%%%
%%%%%%%%%%%%%%%%%%%%%%%%%%%%%%%%%%%%%%%%%%%%%%%%%%%%%%%%%%%%%%%%%%%%%%%%%%%%%%%

\begin{table}[!t]
    \centering
    \includegraphics[scale=0.115]{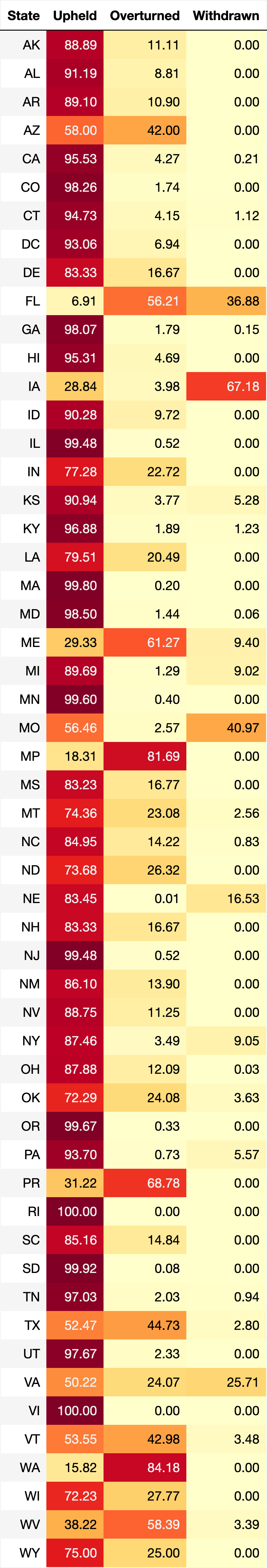}
    \caption{Outcome distribution by state.}
    \label{tab:outcome_by_state}
\end{table}

\paragraph{Distribution of challenge outcomes by state.}
Table~\ref{tab:outcome_by_state} presents the rate at which challenges were upheld, overturned, or withdrawn for each state.
Most states got the majority of their challenges upheld.
Ten had at least 98\% of their challenges upheld.
21 states had at least 90\% upheld.
34 states had at least 80\% upheld.
Six state were exceptions where most challenges were not upheld, Florida, Iowa, Maine, Puerto Rico, Washington, and West Virginia.
These six states each had at least 6700 challenges submitted.
Florida only had about 16,000 or 7\% out of more than 232,000 of its challenges upheld.
Iowa, Missouri, Florida, and Virginia all had a significant percentage of withdrawn challenges.
Among those, Iowa had most (about \num{30000}) of its challenges withdrawn.

%%%%%%%%%%%%%%%%%%%%%%%%%%%%%%%%%%%%%%%%%%%%%%%%%%%%%%%%%%%%%%%%%%%%%%%%%%%%%%%
% TECHNOLOGY %%%%%%%%%%%%%%%%%%%%%%%%%%%%%%%%%%%%%%%%%%%%%%%%%%%%%%%%%%%%%%%%%%
%%%%%%%%%%%%%%%%%%%%%%%%%%%%%%%%%%%%%%%%%%%%%%%%%%%%%%%%%%%%%%%%%%%%%%%%%%%%%%%

\begin{table}[!t]
    \centering
    \includegraphics[scale=0.115]{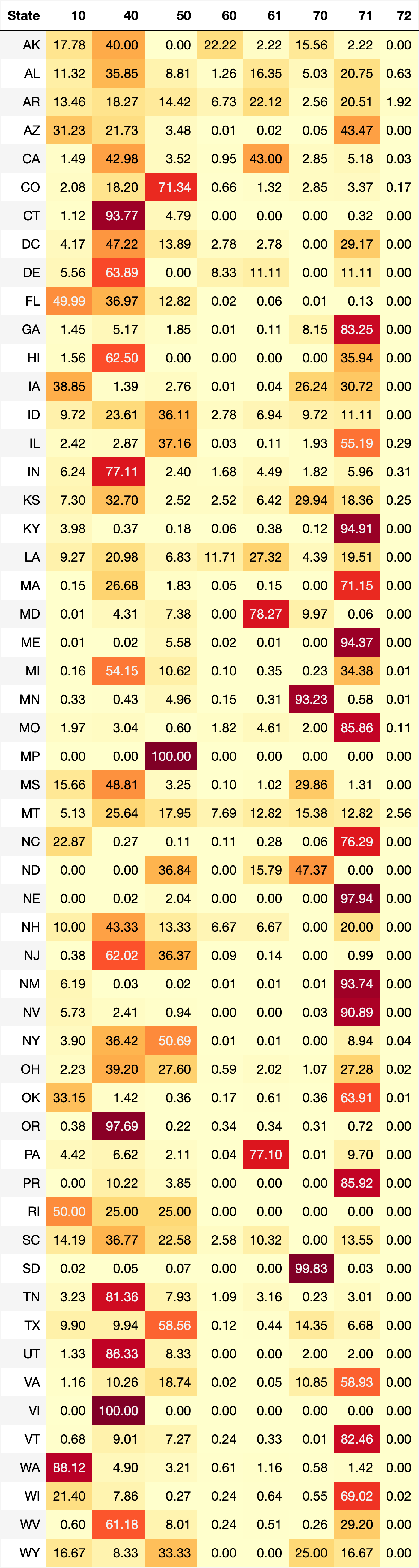}
    \caption{Technology distribution by state.}
    \label{tab:technology_by_state}
\end{table}

\paragraph{Distribution of challenges across technologies by state.}
Tab.~\ref{tab:technology_by_state} presents for each state the rate at which challenges targeted distinct Internet access technologies.
Fifteen states focused the majority of their challenges on \ltfwl availability.
Ten other states focused on \cable availability.
The states of Colorado, Texas and New York had the majority of their challenges submitted against \fiber.
Washing and Rhode Island focused on \copper availability, while Maryland and Pennsylvania focused on \ngssat and Minnesota and South Dakota focused \utfwl.
No state had more than 2\% of challenges against \lbrtfwl, which can be explained by its limited availability at the state-level.
% TODO \input{floats/state/fig_p_60_GS Sat_state.tex}

%%%%%%%%%%%%%%%%%%%%%%%%%%%%%%%%%%%%%%%%%%%%%%%%%%%%%%%%%%%%%%%%%%%%%%%%%%%%%%%
% CATEGORY %%%%%%%%%%%%%%%%%%%%%%%%%%%%%%%%%%%%%%%%%%%%%%%%%%%%%%%%%%%%%%%%%%%%
%%%%%%%%%%%%%%%%%%%%%%%%%%%%%%%%%%%%%%%%%%%%%%%%%%%%%%%%%%%%%%%%%%%%%%%%%%%%%%%

\begin{table}[!t]
    \centering
    \includegraphics[scale=0.115]{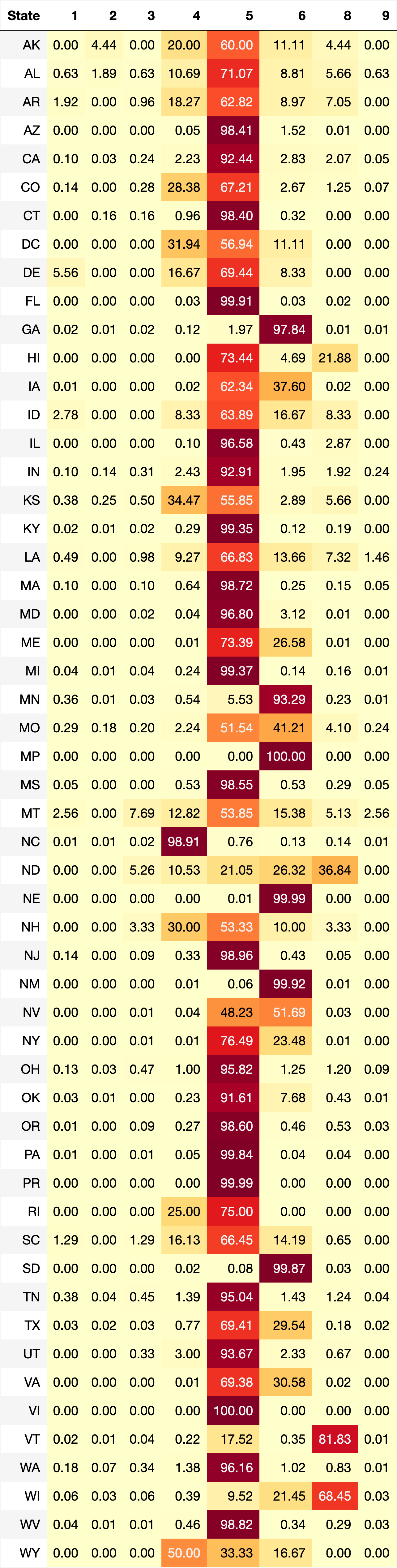}
    \caption{Category distribution by state.}
    \label{tab:category_by_state}
\end{table}

\paragraph{Distribution of challenges across categories by state.}
Tab.~\ref{tab:category_by_state} presents for each state the rate at which challenges were submitted due to each category/reason.
The major category across most (forty one) states was \notechnologyabbr, when a provider does not offer the technology reported to be available at the BSL.
Another six states had \nospeedabbr, when a provider does not offer the speed(s) reported to be available at a location, as the most frequent category.
North Carolina submitted about 99\% of its more than \num{43000} challenges due to \servicedeniedabbr, providers denying requests for service.
Vermont and Wisconsin had most of their challenges due to a wireless or satellite signal not being available at BSLs.
All the other category/reasons accounted for at most about 8\% of all challenges submitted for any state.

\section{Conclusion and Future Work} \label{sec:conclusion}

The analysis carried out so far already brings many interesting insights into the BDC challenge process and its outcome. In the future, we plan to extend this study to consider demographics indicators to answer questions related to how race, income, education, age, and rurality (for example) relate to engagement with the challenge process. We also plan to consider smaller geographic units to understand how challenges are distributed across counties and census tracts, for example. Finally, we will also introduce spatial aspects to the analysis, which could answer questions such as are certain types of challenges clustered in certain areas?

\bibliographystyle{ACM-Reference-Format}
\bibliography{paper}
\clearpage

\end{document}